\documentstyle[12pt]{article}
\begin{document}
\begin{center}
{\Large\bf Mean Charge of the Light Sea Quarks in
the Proton}\vspace{1cm}\\
Susumu Koretune \\
Department of Physics , Fukui Medical University , \\
Matsuoka,  Fukui 910-1193 , Japan \vspace{2cm}\\
\end{center}
The modified Gottfried sum rule multiplied by $3/2$ 
can be interpreted as the one to measure the mean $I_3$
of the [(quark) $-$ (antiquark)] in the proton .  Based on this 
interpretation we find the sum rule which can be understood as
the one to measure the mean charge of the light 
sea quarks (u,d,s) in the proton , and show that it takes the
value $0.23$ for the proton and $0.34$ for the neutron .
\newpage

The Gottfried sum rule \cite{Got} is usually interpreted as
the difference between the charge square of the quarks in the
proton and in the neutron.   
Many years ago,  however,  this sum rule
was related to the bilocal quantity, \cite{MGot84} and several years ago this quantity
was related to the kaon-nucleon scatterings. \cite{MGot93,MGot95} I
called this sum rule 'the modified Gottfried sum rule'.   It takes the form;
\begin{eqnarray}
\lefteqn{\int^1_0\frac{dx}{x}\{F_2^{ep}(x,Q^2)-F_2^{en}(x,Q^2)\}}&&\nonumber \\
&=&\frac{1}{3}\left( 1-\frac{4f_{K}^2}{\pi}\int_{m_Km_N}^{\infty}\frac{d\nu}{\nu^2}
\sqrt{\nu^2-(m_Km_N)^2}\{\sigma^{K^+n}(\nu)-\sigma^{K^+p}(\nu)\}\right) ,
\end{eqnarray}
where $\sigma^{K^+N}(\nu)$ is the total cross section of the 
$K^+N$ scatterings and $f_K$ is the kaon decay constant.  Through 
the experimental values of these quantities,  the right-hand
side of this sum rule was estimated as $0.26\pm 0.03$.   
Equation(1) is derived essentially from the fact that both sides 
of Eq. (1) are related to the quantity
\begin{equation}
\frac{1}{3\pi}P\int_{-\infty}^{\infty}\frac{d\alpha}{\alpha}
A_3(\alpha ,0),
\end{equation}
where $A_3(\alpha ,0)$ is a bilocal quantity which we will explain
soon.  The left-hand side of Eq. (1) multiplied by $3/2$ can be 
expressed in the parton model as
\begin{eqnarray}
\lefteqn{\int_0^1dx\left\{\frac{1}{2}u_v - \frac{1}{2}d_v\right\} +
\int_0^1dx\left\{\frac{1}{2}\lambda_u - \frac{1}{2}\lambda_d\right\}
- \int_0^1dx\left\{- \frac{1}{2}\lambda_{\bar{u}} + 
\frac{1}{2}\lambda_{\bar{d}}\right\}}&& \nonumber \\
& =& \frac{1}{2} 
+\frac{1}{2}\int_0^1dx \{\lambda_u - \lambda_d
+\lambda_{\bar{u}}- \lambda_{\bar{d}}\}\hspace*{5cm}
\end{eqnarray}
where $\lambda_i$ is the sea quark of the {\it i} quark.    
Thus we can understand it as the mean $I_3$ of the 
[(quark) $-$ (antiquark)] in the proton.  This agrees with Eq. (2)
in the sense that it has the meanings of the mean $I_3$ of something.      
Though the bilocal quantity in our formalism is not
necessarily that defined by the quark field,  we can give
clear correspondence between them.\cite{Ant81}    
As far as the $n=1$ moment of the structure
function $F_2$ is concerned,    
$A_a(\alpha ,0)$ can be considered as the quantity defined by
\begin{equation}
\left\langle p|\frac{1}{2i}\left[\bar{q}(x)\gamma^{\mu}\frac{1}{2}\lambda_a q(0)
-\bar{q}(0)\gamma^{\mu}\frac{1}{2}\lambda_a q(x)\right]|p\right\rangle_c
=p^{\mu}A_a(px,x^2)+x^{\mu}\bar{A}_a(px,x^2),
\end{equation}
on the null-plane $x^+=0$.  It should be noted that as a particular
property of the method to reach the fixed-mass sum rule in the
null-plane formalism the state $|p\rangle$ can be taken in any frame.
\cite{DJT}   Thus we can even take the rest frame which may be useful
in low energy models such as chiral quark models and
soliton models.   Now we can see why the contribution from the
antiquark is multiplied by $-1$ in Eq. (2).  Decomposing the
quark field into the particle mode and the anti-particle one , 
we find that because of the integral of the type
$P\int_{-\infty}^{\infty}\frac{d\alpha}{\alpha}\dots$,
the contribution from the anti-particle mode gets an
additional factor $-1$. 
Hence the sea quark and its antiquark
contribute additively to the sum rule. 
Compared with this,the Adler sum rule \cite{Adler} corresponds
to the mean $I_3$ of the [(quark)$+$(antiquark)] in the proton. 
Hence the contribution
from the sea quark and its antiquark to the sum rule 
cancels out exactly, and it measures the mean $I_3$ of  
the valence quark being equal to the $I_3$ of the
proton.  This is the fundamental difference
between the Gottfried sum rule and the Adler sum rule.
The experimental study starting from the Ellis-Jaffe sum rule 
\cite{Ellis,Ellisexp} and the Gottfried sum rule \cite{Gotexp} 
in recent ten years has shown that the hadronic vacuum
is very important in understanding the deep structure of the 
hadron.   HERA \cite{HERA} also has shown that the theoretical 
understanding of the pomeron which should be related to this 
hadronic vacuum is very important.   
In view of these situations it is 
important to have a model-independent constraint on  
the hadronic vacuum which has a clear physical meaning.    
Here we give the sum rule which can be understood as the mean 
charge of the light sea quarks in the proton, where the 'light 
sea quarks' are those of $u,$ $d,$ and $s$ types.

Let us first derive the hypercharge sum rule in the
$SU(3)$ flavor group .   Explicit forms of the various sum rules
in our formalism including their derivation are reviewed 
in Ref. \cite{MGot95}. According to this,  it is straightforward 
to obtain the sum rule
\begin{eqnarray}
\lefteqn{\frac{1}{2\pi}\frac{2\sqrt{3}}{3}P\int_{-\infty}^{\infty}\frac{d\alpha}{\alpha}
A_8(\alpha ,0)}\nonumber \\
&=&\int_0^1\frac{dx}{x}\{F_2^{\bar{\nu}p}(x,Q^2) +
F_2^{\nu p}(x,Q^2) - 3F_2^{ep}(x,Q^2) - 3F_2^{en}(x,Q^2)\} ,
\end{eqnarray}
\begin{equation}
\frac{1}{2\pi}\frac{2\sqrt{3}}{3}P\int_{-\infty}^{\infty}\frac{d\alpha}{\alpha}
A_8(\alpha ,0)=\frac{1}{3}[2I_{\pi} - I_K^p - I_K^n] ,
\end{equation}
where $I_{\pi},I_K^p$ and $I_K^n$ are defined in Ref. \cite{MGot95} by assuming the
smooth extrapolation to the on-shell quantity as
\begin{eqnarray}
I_{\pi}=\lefteqn{g_A^2(0) +\frac{2f_{\pi}^2}{\pi}
\int_{\nu_0^{\pi}}^{\infty}\frac{d\nu}{\nu^2}[(\nu^2-m_{\pi}^2m_{N}^2)^{1/2}
\{\sigma^{\pi^+p}(\nu)+\sigma^{\pi^-p}(\nu)\}-\nu s^{b}\beta_{\pi
  N}]}\nonumber \\
&&+\frac{2f_{\pi}^2\beta_{\pi N}}{\pi}\ln (\frac{1}{2\nu_0^{\pi}}) \;.\hspace*{9cm} 
\end{eqnarray}
\begin{eqnarray}
I_{K}^p=\lefteqn{(g_A^{p\Sigma^0}(0))^2 +(g_A^{p\Lambda^0}(0))^2 + \frac{2f_{K}^2}{\pi}
\int_{\nu_0^{K}}^{\infty}\frac{d\nu}{\nu^2}[(\nu^2-m_{K}^2m_{N}^2)^{1/2}}&&  \nonumber \\
&&\{\sigma^{K^+p}(\nu)+\sigma^{K^-p}(\nu)\} -\nu s^{b}\beta_{KN}] 
+\frac{2f_{K}^2\beta_{KN}}{\pi}\ln (\frac{1}{2\nu_0^{K}}) + U_p \;.\hspace*{2cm}
\end{eqnarray}
\begin{eqnarray}
I_{K}^n=\lefteqn{(g_A^{n\Sigma^-}(0))^2  + \frac{2f_{K}^2}{\pi}
\int_{\nu_0^{K}}^{\infty}\frac{d\nu}{\nu^2}[(\nu^2-m_{K}^2m_{N}^2)^{1/2}
\{\sigma^{K^+n}(\nu)+\sigma^{K^-n}(\nu)\}}&&  \nonumber \\
&&-\nu s^{b}\beta_{KN}]
+\frac{2f_{K}^2\beta_{KN}}{\pi}\ln (\frac{1}{2\nu_0^{K}}) + U_n \;.\hspace*{6cm}
\end{eqnarray}
Here the intercept of the pomeron is assumed as
$\alpha_P(0)=1+b$ with $b=0.0808$,\cite{Land} and $s$ is defined as
$s=m_{\pi}^2+m_N^2+2\nu$ for $I_{\pi}$ and
$s=m_{K}^2+m_N^2+2\nu$ for $I_K^p$ and $I_K^n$. The quantities $\beta_{\pi N}$ and
$\beta_{KN}$ correspond to the residues of the pomeron which subtract
the infinity in the above each integral.  The quantities $\nu_0^{\pi}$
and $\nu_0^{K}$ are defined 
as $\nu_0^{\pi}=m_{\pi}m_N$ and $\nu_0^{K}=m_{K}m_N$.  
The terms $U_p$ and $U_n$ are the contributions below the $\bar{K}N$ 
threshold.   Using Adler-Weisberger sum rules for the kaon,\cite{AdlerWeis}  
we can express these terms by the integral over the $KN$ total cross sections. 
Then , through the experimental values of $\pi N$ and $KN$ total cross
sections,  $I_{\pi},I_K^p$ and $I_K^n$
were estimated as $I_{\pi}\sim 5.17$, $I_K^p\sim 2.39$  
and $I_K^n\sim 1.61$.\cite{MGot95}  Thus the right-hand side
of the sum rule (6) is 
$\frac{1}{3}[2I_{\pi} - I_K^p - I_K^n]\sim 2.12$.   
Subtracting the contribution from the valence quarks from this,  we find that 
the mean hypercharge of the light sea quarks is 
$(2.12-1)/2\sim 0.56$.  The reason $(2.12-1)$ is divided by $2$ is 
that the sea quarks and their antiquarks contribute additively.   
Similarly, the mean $I_3$ of the light sea quarks given by the modified 
Gottfried sum rule is $3(0.26-1/3)/4\sim -0.055$.   Thus we obtain the sum rule 
of the mean charge of the light sea quarks in the proton and its value as
\begin{eqnarray}
\lefteqn{<Q>_{{\rm light\;sea\;quarks}}^{{\rm proton}}}&&\nonumber\\
&=&\frac{1}{2}\left[ \left\{\frac{1}{2\pi}P
\int_{-\infty}^{\infty}
\frac{d\alpha}{\alpha}A_3(\alpha ,0)-\frac{1}{2}\right\} + \frac{1}{2}\left\{
\frac{1}{2\pi}\frac{2\sqrt{3}}{3}P\int_{-\infty}^{\infty}\frac{d\alpha}{\alpha}
A_8(\alpha ,0) -1 \right\} \right]  \nonumber \\
&=&\frac{1}{6}(I_{\pi} + I_K^p - 2I_K^n)-\frac{1}{2} \sim 0.23 .\hspace{5.9cm}
\end{eqnarray}
Because of the large positive mean hypercharge,  the mean charge 
becomes positive, though the mean $I_3$ is negative.   It goes without saying 
that the mean charge of the light antiquarks in the proton is 
$<Q>_{{\rm light\;antiquarks}}^{{\rm proton}}\sim -0.23$.   
In the neutron case we find 
\begin{eqnarray}
\lefteqn{<Q>_{{\rm light\;sea\;quarks}}^{{\rm neutron}}}&&\nonumber \\
&=&\frac{1}{2}\left[ -\left\{ \frac{1}{2\pi}P
\int_{-\infty}^{\infty}
\frac{d\alpha}{\alpha}A_3(\alpha ,0)-\frac{1}{2}\right\} + \frac{1}{2}\left\{
\frac{1}{2\pi}\frac{2\sqrt{3}}{3}P\int_{-\infty}^{\infty}\frac{d\alpha}{\alpha}
A_8(\alpha ,0) -1 \right\} \right]   \nonumber \\
&=&\frac{1}{6}(I_{\pi} - 2I_K^p + I_K^n) \sim 0.34 .\hspace{6.9cm}
\end{eqnarray}
Since the right-hand side of Eq. (1) is $(I_K^p-I_K^n)/3$ ,
from Eqs. (1),(2) and (5),  we can easily recognize 
that Eqs. (10) and (11) have the correct meaning of the mean charge
of the light sea quarks in the proton in the parton model.  
However, these results do not depend on the 
parton model.  They give us a model independent constraint on 
the matrix elements of the bilocal currents in 
an arbitrary frame of the nucleon.

The perturbative QCD corrections to the relations
(6),(10) and (11) begin from 2 loops in the anomalous
dimension and they enter in the same way as that in
the Gottfried sum rule. \cite{Ross}  Therefore they
are negligible compared with the non-perturbative values
given in this paper.
Further, it is possible to check Eqs. (10) and (11)
by the parton distributions available at present, such as
the ones by the CTEQ group \cite{CTEQ} or the MRS group. \cite{MRS}
Before the numerical calculation , however,
we can recognize that these distributions will not satisfy
Eqs. (10) and (11), since they do not satisfy the 
symmetry constraint $\lim_{x \to 0}x^{\alpha_P(0)}\lambda_u(x,Q^2)=
\lim_{x \to 0}x^{\alpha_P(0)}\lambda_d(x,Q^2)=
\lim_{x \to 0 }x^{\alpha_P(0)}\lambda_s(x,Q^2)$,
which comes in our approach from the fact that the pomeron
is flavor singlet, and since Eqs. (10) and (11)
depend heavily on the behavior in the small $x$ region.
To see this fact more concretely, we give here a typical
example by the CTEQ4M initial set at $Q_0 = 1.6$GeV.
By changing the lower limit of the $x$ integration
to $1\times 10^{-6}$, $1\times 10^{-5}$, $1\times 10^{-4}$
and $1\times 10^{-3}$, we find that the contribution
above $x=1\times 10^{-3}$ to Eq. (10) is 0.24, and hence it already
satisfies the relation.  However the contribution 
in the region $1\times 10^{-4}\leq x \leq1\times 10^{-3}$ 
is 0.17, $1\times 10^{-5}\leq x \leq1\times 10^{-4}$ is 0.22,
and $1\times 10^{-6}\leq x \leq1\times 10^{-5}$ is 0.30.
Thus we see that the large contribution comes from the region
below $x=1\times 10^{-3}$. The origin of this is clear.
The strange sea quark is too small in comparison with the up and 
down sea quarks, while the up and down sea quarks are
very similar in this region. Since the sea quarks
in the small $x$ region are poorly determined in a
phenomenological analysis, we see that the relations
(10) and (11) are helpful also in this respect .

In conclusion we derived a sum rule
which can be understood in the parton model as that to
measure the mean charge of the light sea quarks 
in the proton and show that this charge takes the value
$0.23(0.34)$ for the proton(neutron).

\end{document}